\documentclass[a4paper,conference]{IEEEtran}
\IEEEoverridecommandlockouts

\usepackage[letterpaper,top=0.75in,bottom=1.035in,left=0.635in,right=0.635in]{geometry}
\usepackage{cite}
\usepackage{amsmath,amssymb,amsfonts}
\usepackage{graphicx}
\usepackage{textcomp}
\usepackage{xcolor}
\usepackage[nolist]{acronym}
\usepackage{mathtools} 
\usepackage{graphicx}
\usepackage{textcomp}
\usepackage{xcolor}
\usepackage{tabularx}
\usepackage{booktabs}
\usepackage{float}
\usepackage{algpseudocode}
\usepackage[english]{babel}
\usepackage{graphicx,lipsum}
\usepackage{paralist}
\usepackage{array}
\usepackage{amsmath,stackrel}
\usepackage{subfig}
\usepackage{multicol}
\usepackage{verbatim}
\usepackage{enumerate}              
\usepackage{dsfont}
\usepackage{psfrag}
\usepackage{algorithm}
\usepackage[nolist]{acronym}
\usepackage[utf8]{inputenc}
\usepackage{tikz}
\usepackage{collcell}
\usepackage{pgfplots}
\pgfplotsset{compat=1.18}
\usepackage{hhline}
\usepackage{colortbl}
\usepackage{multido}
\usepackage{multirow}
\usepackage{dblfloatfix}    
\usepackage{etoolbox}
\usepackage{soul}
\usepackage{calligra}
\usepackage[font=footnotesize]{caption}
\def\BibTeX{{\rm B\kern-.05em{\sc i\kern-.025em b}\kern-.08em
    T\kern-.1667em\lower.7ex\hbox{E}\kern-.125emX}}

\newcommand{\transp}{\mathsf{T}}

\DeclarePairedDelimiter\ceil{\lceil}{\rceil}

\begin{acronym}
\acro{AWGN}{additive white Gaussian noise}
\acro{AoA}{angle of arrival}
\acro{AoD}{angle of departure}
\acro{BF}{beamformer}
\acro{BS}{base station}
\acro{CIR}{channel impulse response}
\acro{CP}{cyclic prefix}
\acro{CKF}{cubature Kalman filter}
\acro{CPHD}{cardinalized probability hypothesis density}
\acro{DFT}{discrete Fourier transform}
\acro{DBSCAN}{density-based spatial clustering of applications with noise}
\acro{DoA}{direction of arrival}
\acro{DoD}{direction of departure}
\acro{EIRP}{effective isotropic radiated power}
\acro{ELP}{equivalent low-pass}
\acro{ESPRIT}{Estimation of Signal Parameters via Rotational Invariance Techniques}
\acro{FC}{fusion center}
\acro{FD}{frequency division}
\acro{FFT}{fast Fourier transform}
\acro{FDD}{frequency-division duplexing}
\acro{FAR}{false alarm rate}
\acro{GM}{Gaussian mixture}
\acro{GMPHD}{Gaussian mixture probability hypothesis density}
\acro{GMCPHD}{Gaussian mixture cardinalized probability hypothesis density}
\acro{GOSPA}{generalized optimal subpattern assignment}
\acro{ICI}{inter-carrier interference}
\acro{InF}{indoor factory}
\acro{IoT}{internet of things}
\acro{IFFT}{inverse fast Fourier transform}
\acro{ISI}{inter-symbol interference}
\acro{i.i.d.}{independent, identically distributed}
\acro{JSC}{joint sensing and communication}
\acro{k-NN}{k-nearest neighbors}
\acro{LOS}{line-of-sight}
\acro{LTE}{long term evolution}
\acro{MAP}{maximum a posteriori}
\acro{MB}{multi-Bernoulli}
\acro{MBM}{multi-Bernoulli mixture}
\acro{MCL}{maximum coupling loss}
\acro{MPL}{maximum path loss}
\acro{MIL}{maximum isotropic loss}
\acro{MC}{Monte Carlo}
\acro{MIMO}{multiple-input multiple-output}
\acro{MUSIC}{MUltiple SIgnal Classification}
\acro{MDL}{minimum description length}
\acro{MHT}{multi-hypothesis tracker}
\acro{mmWave}{millimeter-wave}
\acro{NR}{new radio}
\acro{NLoS}{non-line-of-sight}
\acro{OFDM}{orthogonal frequency division multiplexing}
\acro{OSPA}{optimal sub-pattern assignment}
\acro{p.d.f.}{probability density function}
\acro{PF}{particle filter}
\acro{PHD}{probability hypothesis density}
\acro{PSD}{power spectral density}
\acro{PPP}{Poisson point process}
\acro{PSK}{phase shift keying}
\acro{QAM}{quadrature amplitude modulation}
\acro{QPSK}{quadrature phase shift keying}
\acro{QuaDRiGa}{QUAsi Deterministic RadIo channel GenerAtor}
\acro{RCS}{radar cross-section}
\acro{RF}{radio frequency}
\acro{RFS}{random finite set}
\acro{RMSE}{root mean squared error}
\acro{r.v.}{random variable}
\acro{RRH}{remote radio head}
\acro{Rx}{receiver}
\acro{SSIR}{signal-to-self interference ratio}
\acro{SI}{self interference}
\acro{SNR}{signal-to-noise ratio}
\acro{SU}{single-user}
\acro{SCM}{sample covariance matrix}
\acro{SPF}{soft particle filter}
\acro{TD}{time-division}
\acro{TDD}{time-division duplexing}
\acro{ToA}{time of arrival}
\acro{THz}{terahertz}
\acro{TDoA}{time difference of arrival}
\acro{Tx}{transmitter}
\acro{UMi}{urban micro}
\acro{UE}{user equipment}
\acro{ULA}{uniform linear array}
\acro{V2V}{vehicle-to-vehicle}
\end{acronym}

\begin{document}

\title{Multi-Target Acquisition in Multistatic MIMO-OFDM Joint Sensing and Communication\\

}

\author{
Elisabetta~Matricardi,
Lorenzo~Pucci, 
Elia~Favarelli,
Enrico~Paolini, 
and Andrea~Giorgetti
\thanks{Authors are with the Wireless Communications Laboratory (WiLab), CNIT, and DEI, University of Bologna, Italy.
Email: \{elisabett.matricard3, lorenzo.pucci3, elia.favarelli, e.paolini, andrea.giorgetti\}@unibo.it}
\thanks{This work was supported in part by the CNIT National Laboratory WiLab and the WiLab-Huawei Joint Innovation Center and in part by the European Union under the Italian National Recovery and Resilience Plan (NRRP) of NextGenerationEU, partnership on ``Telecommunications of the Future'' (PE00000001 - program ``RESTART'').}
}

\maketitle

\begin{abstract}
In this work, we investigate a multistatic MIMO-OFDM \ac{JSC} system that leverages cooperation among spatially distributed \acp{BS} to detect and localize multiple targets through soft fusion of range-angle maps. We propose an innovative selective data fusion strategy that combines only the most reliable regions of range-angle maps from each bistatic pair, mitigating the adverse effects of residual clutter and target smearing inherent to bistatic configurations. To further enhance multi-view perception, we introduce a round-robin transmitter role strategy, enabling \acp{BS} to cooperate and exploit target spatial diversity. Finally, we assess the system performance in a cluttered environment using the \ac{GOSPA} and \ac{RMSE} metrics, demonstrating the effectiveness of our approaches in improving detection and localization. 
\end{abstract}


\acresetall
\section{Introduction}\label{sec:intro}
The \ac{JSC} paradigm is emerging as a cornerstone of 6G, helping in improving spectral efficiency, reducing latency and energy consumption, and lowering hardware complexity and costs while enabling sensing for scenarios such as autonomous vehicles, smart cities, and public safety monitoring. Furthermore, advances in \ac{mmWave}, \ac{THz}, and massive \ac{MIMO} technologies have enabled high-accuracy sensing using communication signals \cite{LiXiaoZheng:24}, seamlessly integrating sensing within cellular networks. 
In such a scenario, multistatic radar supported by advanced network architectures like Cloud-RAN leverages spatially distributed \acp{Tx} and \acp{Rx} to enhance detection and coverage, particularly in cluttered urban environments while avoiding the full-duplex technology required by monostatic systems \cite{FullDuplex}. 
Processing \ac{ToA}, \ac{AoA}, or \ac{TDoA} estimates at individual sensors followed by data fusion is a standard approach but is susceptible to missed detections, particularly for weak targets. Additionally, in multi-target scenarios, a data association step is required, which can be computationally complex \cite{Biruk:11}.

In contrast, soft data fusion integrates target echoes from all sensors at the \ac{FC}, providing a more comprehensive view \cite{Godrich:10, Matricardi:23}. Although this approach entails higher processing and signaling overhead, it preserves information from all sensors, significantly enhancing detection accuracy. To the authors' knowledge, few studies have explored multistatic \ac{JSC} systems with an \ac{FC} for multi-target detection and parameter estimation via soft data fusion, leveraging MIMO-OFDM technology \cite{Rappaport:21, Tagliaferri:J24}.

To address this gap, this work investigates sensing in a multistatic \ac{MIMO}-\ac{OFDM} \ac{JSC} system operating at \ac{mmWave} frequencies, with a focus on the target acquisition phase. Specifically, we analyze a multistatic \ac{JSC} system where multiple point-like targets are detected and localized by illuminating the scene and fusing range-angle maps from multiple bistatic pairs. In each radar measurement, only one \ac{Tx} is active, while the remaining \acp{BS} serve as \acp{Rx}.

A key aspect of this work is the adoption of a \emph{selective data fusion} strategy, where only reliable regions of each range-angle map are used to enhance cooperation and performance. Additionally, we introduce a \emph{round-robin mechanism} in which \acp{BS} take turns acting as \acp{Tx}, enabling multi-view perception. A fused radar map is generated once all \acp{BS} have participated as \acp{Tx}. Numerical simulations demonstrate that the proposed selective data fusion approach and round-robin mechanism achieve high localization accuracy and detection performance in multi-target scenarios.
%

In this paper, bold uppercase and lowercase letters represent matrices and vectors, respectively, $[\mathbf{X}]_{a,b}$ denotes the element $(a,b)$ of a matrix $\mathbf{X}$, while $(\cdot)^\transp$, $(\cdot)^\mathsf{c}$, and $\|\cdot\|_p$ denote the transpose, conjugate, and p-norm operators, respectively. Moreover, $\mathbb{E}\{\cdot\}$ denotes the expected value, $\mathbf{n} \sim \mathcal{CN}(\mathbf{0}, \boldsymbol{\Sigma})$ represents a zero-mean circularly symmetric complex Gaussian vector with covariance $\boldsymbol{\Sigma}$, and $\mathbf{I}_N$ is the $N \times N$ identity matrix. The operator $|\cdot|$ denotes either the absolute value function or the cardinality of a set, depending on the context, $\odot$ is the element-wise product, and  $\ceil{\cdot}$ represents the ceiling function.

The rest of the paper is organized as follows. Section~\ref{sec:sysmodel} presents the system model, Section~\ref{sec:search} presents range angle-map calculation and soft data fusion, Section~\ref{sec:results} provides numerical results, and Section~\ref{sec:conclu} concludes the paper with remarks.

\section{System Model}\label{sec:sysmodel}
\begin{figure}
    \centering
\includegraphics[width=0.9\linewidth]{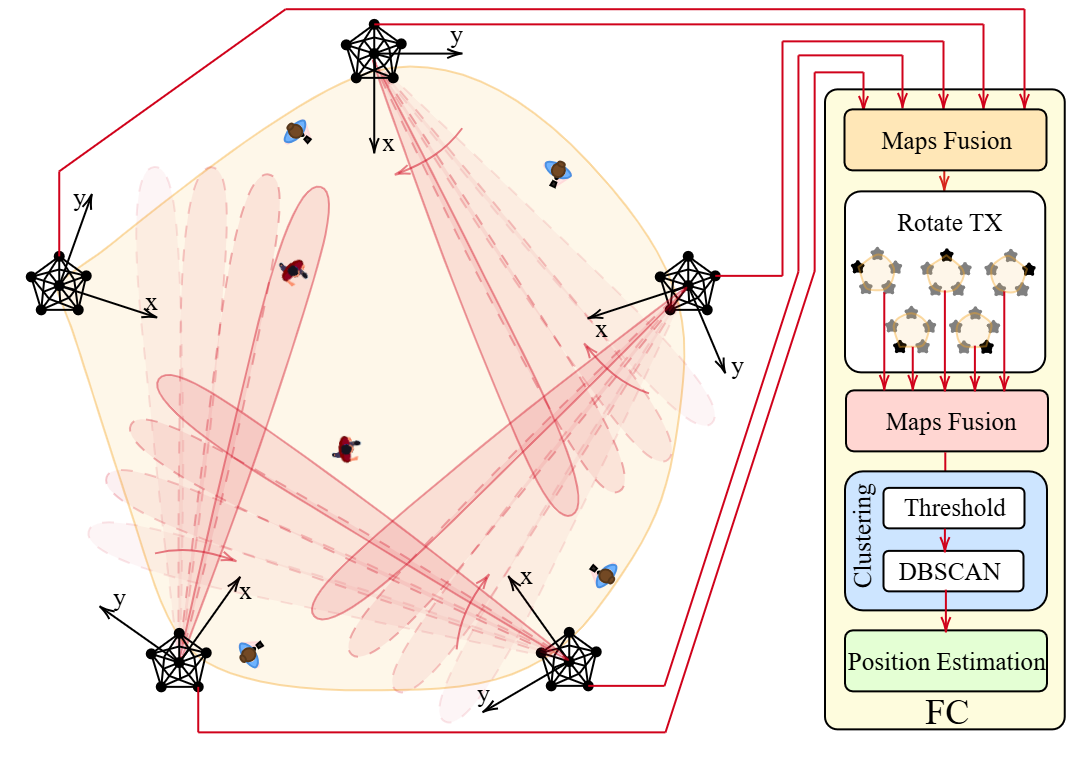}
    \caption{The multistatic system where a wide Tx beam illuminates the area of interest, and Rx beams scan to detect targets.}
    \label{fig:search}
\end{figure}

We consider a multistatic \ac{JSC} system, depicted in Fig.~\ref{fig:search}, where sensing is performed during the downlink communication between a \ac{BS}, serving as the \ac{Tx}, and \acp{UE}, where the remaining \acp{BS} act as sensors in receiving mode. Let us define the set $\mathcal{S} = \{\mathbf{s}_1, \mathbf{s}_2, \dots, \mathbf{s}_{|\mathcal{S}|}\}$, 
which contains the position $\mathbf{s}_i = (s_{\mathrm{x},i}, s_{\mathrm{y},i})$ of all the \acp{BS}, being $|\mathcal{S}|$ the total number of \acp{BS}. More precisely, we write $\mathcal{S}_t$ to
indicate that the $t$th \ac{BS} acts as the \ac{Tx}, subject to change over time. Without loss of generality, the \acp{BS} are equipped with two half-wavelength spaced \acp{ULA} of $N_\mathrm{T}$ elements and $N_\mathrm{R}$ elements as \ac{Tx} and \ac{Rx}, respectively. 
Each \ac{BS} collects signals reflected from targets, pre-processes such signals locally, and transmits processed data to an \ac{FC} at the network's edge; the \ac{FC} may be part of a Cloud-RAN.

The network aims to detect and estimate target positions within the coverage area without prior knowledge of their number or location. Let $\boldsymbol{\mathcal{Z}} = \{\mathbf{z}_1, \mathbf{z}_2, \dots, \mathbf{z}_Q\}$ represent the positions of $Q$ targets in the area, where $\mathbf{z}_q = (z_{\mathrm{x},q}, z_{\mathrm{y},q})$. For a multistatic configuration $\mathcal{S}_t$, comprising $|\mathcal{S}|-1$ bistatic pairs with the same \ac{Tx}, each target lies on an ellipse with foci at the \ac{Tx} and \ac{Rx} positions, with major axis corresponding to the bistatic range \cite{Pucci:22Bistatic}. For each target $q$, the bistatic range for the \ac{Tx}-\ac{Rx} pair $(t,i)$ is $R_{t,q,i} = r_{t,q} + r_{q,i} = c \, \tau_{t,q,i}$, where $r_{t,q}$ and $r_{q,i}$ are the distances from the \ac{Tx} and \ac{Rx} to the target, respectively, $\tau_{t,q,i}$ is the \ac{ToA}, and $c$ the speed of light. The distance between \ac{Tx} $t$ and \ac{Rx} $i$ is the baseline $L_{t,i}$ \cite{willis05}.
To avoid \ac{ISI}, the channel delay spread, measured from the \ac{LOS} path component with \ac{ToA} $\tau_{t,i} = L_{t,i}/c$, must be less than the guard time $T_\mathrm{g}$ between \ac{OFDM} symbols. As a result, the maximum detectable bistatic range is constrained by $R_{\mathrm{bis},t,i}^\mathrm{max} < c \, T_\mathrm{g} + L_{t,i}$. This limits the observable region for bistatic pair $(t,i)$ to an ellipse with a minor axis of length $\sqrt{(T_\mathrm{g}c+L_{t,i})^2 - L_{t,i}^2}$.
\subsection{Transmitted and Received Signals}\label{sec:in-out-rel}
The transmitted signal consists of $M$ \ac{OFDM} symbols with $K$ active subcarriers, forming a $K \times M$ matrix of modulation symbols $x_{k,m}$, each normalized so that $\mathbb{E}\{|x_{k,m}|^2\}=1$.\footnote{For the sake of simplicity, hereinafter, a generic \ac{Tx} is considered and the index $t$ is omitted. This is replaced by a generic $\mathrm{T}$ only when necessary.} At the \ac{Tx}, each symbol is mapped onto the \ac{Tx} antenna array as $\tilde{\mathbf{x}}[k,m] = \mathbf{w}_\mathrm{T} x_{k,m} \in \mathbb{C}^{N_\mathrm{T} \times 1}$, where $\mathbf{w}_\mathrm{T}$ is the beamforming vector such that $\|\mathbf{w}_\mathrm{T}\|^2_2=P_\mathrm{avg}$, being $P_\mathrm{avg}=P_\mathrm{T}/K$ the transmit power per subcarrier and $P_\mathrm{T}$ the total transmit power. In this work, $\mathbf{w}_\mathrm{T}$ is properly designed to generate a radiation pattern with nearly uniform gain on a wide circular sector \cite{Friedlander:12}. This large, uniform beam is crucial since the objective is to illuminate simultaneously all potential targets within the monitored area, without needing beam alignment between \ac{Tx} and \acp{Rx}. 
After modulation through \ac{IFFT} and up-conversion, the precoded symbols are transmitted through the wireless channel. 
Assuming negligible \ac{ISI} and \ac{ICI}, after down-conversion and \ac{FFT}, the $N_\mathrm{R} \times 1$ vector of received symbols at the $i$th \ac{Rx}, subcarrier $k$ and time $m$, is given by
\begin{equation} 
\label{eq:y_tilde}
\tilde{\mathbf{y}}_{i}[k,m] = \mathbf{H}_{i}[k,m] \tilde{\mathbf{x}}[k,m] + \tilde{\mathbf{n}}_{i}[k,m] 
\end{equation}
\noindent where $\tilde{\mathbf{n}}_{i}[k,m]\sim \mathcal{CN}(\mathbf{0}, \sigma_\mathrm{N}^2 \mathbf{I}_{N_\mathrm{R}})$ represents the complex \ac{AWGN} vector, with  $\sigma_\mathrm{N}^2 = N_0 \Delta f$ being $N_0$ the noise \ac{PSD}, while $\mathbf{H}_{i}[k,m] \in \mathbb{C}^{N_\mathrm{R} \times N_\mathrm{T}}$ is the frequency-domain channel matrix between the \ac{Tx} and the \ac{Rx} $i$ for subcarrier $k$ and \ac{OFDM}~symbol $m$. Considering a scenario with $Q$ point-like targets and $\mathcal{L}$ ground clutter points, the channel matrix can be expressed as
%
%
\begin{equation}
\mathbf{H}_{i}[k,m] =  \sum_{\ell = 1}^{Q + \mathcal{L}} \alpha_{\ell,i} 
e^{j2\pi m T_\mathrm{s} f_{\mathrm{d},\ell,i}} 
e^{-j2\pi k \Delta f \tau_{\ell,i}} \mathbf{b}(\theta_{\ell,i}) \mathbf{a}^\mathsf{T}(\phi_{\ell}). 
\label{eq:bistatic_channelMatrix_with_clutter}
\end{equation}
The summation in \eqref{eq:bistatic_channelMatrix_with_clutter} accounts for both target reflections (for $1 \leq \ell \leq Q$) and static ground clutter (for $Q < \ell \leq \mathcal{L}$), where $\alpha_{\ell,i}$ represents the channel gain. Additionally, $\tau_{\ell,i}$ and $f_{\mathrm{d},\ell,i}$ denote the \ac{ToA} and bistatic Doppler shift, respectively. The vectors $\mathbf{a}(\phi_{\ell})$ and $\mathbf{b}(\theta_{\ell,i})$ correspond to the \ac{Tx} and \ac{Rx} array response vectors, where $\phi_{\ell}$ and $\theta_{\ell,i}$ are the direction of departure (DoD) and direction of arrival (DoA), respectively.\footnote{For a \ac{ULA} with $N_\mathrm{a}$ elements spaced half a wavelength apart, as considered in this work, the steering vector for a generic direction $\theta$ is given by:
\begin{equation}
    \label{eq:steeringVec}
    \mathbf{a}(\theta) = \left[ e^{-j\pi \frac{N_\mathrm{a}-1}{2} \sin{\theta}}, \dots, e^{j\pi \frac{N_\mathrm{a}-1}{2} \sin{\theta}} \right]^\mathsf{T}.
\end{equation}}
Here, $\Delta f$ is the subcarrier spacing, and $T_\mathrm{s}=1/\Delta f+T_\mathrm{g}$ is the total \ac{OFDM} symbol duration, including the cyclic prefix $T_\mathrm{g}$. 
Since the clutter is supposed to be static, its bistatic Doppler shift is zero, i.e.,  $f_{\mathrm{d},\ell,i} = 0$ for $Q < \ell \leq \mathcal{L}$. The complex channel gain $\alpha_{\ell,i}$ accounts for attenuation and phase shift along the path between the \ac{Tx}, the scatterer $\ell$, and the \ac{Rx} $i$. Based on the radar equation, its magnitude is
%
%
\begin{equation} 
\label{eq:amplitude}
|\alpha_{\ell,i}|  = \sqrt{\frac{G_\mathrm{R}c^2 \sigma_{\ell,i}}{ (4\pi)^3 f_\mathrm{c}^2(r_{\mathrm{T},\ell}\, r_{\ell,i})^2}} 
\end{equation}
where $f_\mathrm{c}$ is the carrier frequency, $G_\mathrm{R}$ is the single antenna gain, 
and $\sigma_{\ell,i}$ represents the \ac{RCS} of the $\ell$th reflection point. In this work, targets and clutter adhere to the Swerling~I model, where the \ac{RCS} is a \ac{r.v.} that follows an exponential distribution to account for reflection variability \cite{Sko:B08}. Notably, spatially separated \acp{Rx} may observe different target scattering profiles (i.e., \ac{RCS} values), a phenomenon known as spatial diversity, which can be exploited in cooperative target detection \cite{Fishler_spatial_diversity}.
By replacing \eqref{eq:bistatic_channelMatrix_with_clutter} in \eqref{eq:y_tilde} and expanding $\tilde{\mathbf{x}}[k,m]$, the received symbol vector at subcarrier $k$ and time $m$ for the $i$th \ac{Rx} can be rewritten as
\begin{align}\label{eq:y_tilde_final}
\tilde{\mathbf{y}}_{i}&[k,m] = \\
  = & \sum_{\ell = 1}^{Q+\mathcal{L}} x_{k,m}\alpha_{\ell,i} e^{j2\pi m T_\mathrm{s} f_{\mathrm{d},\ell,i}} 
 e^{-j2\pi k \Delta f \tau_{\ell,i}} \mathbf{b}(\theta_{\ell,i})\underbrace{\mathbf{a}^\mathsf{T}(\phi_{\ell}) \mathbf{w}_\mathrm{T}}_{\triangleq \gamma(\phi_\ell)} \nonumber
\end{align}
where $|\gamma(\phi_\ell)|^2 = P_\mathrm{avg} G^{\mathrm{a}}_\mathrm{T}(\phi_\ell)$ represents the \ac{EIRP} towards the $\ell$th scatterer, incorporating the beamforming gain $G^{\mathrm{a}}_\mathrm{T}(\phi_\ell)$, which depends on the considered direction $\phi_\ell$ for a given beamforming vector $\mathbf{w}_\mathrm{T}$. At the $i$th \ac{Rx} digital beamforming is performed using the weight vector $\mathbf{w}_{\mathrm{R},i} \in \mathbb{C}^{N_\mathrm{R} \times 1}$. Details on beamforming design are provided in Section~\ref{sec:range-angle}. Starting from \eqref{eq:y_tilde}, the combined received symbol is then $y_{i,k,m} = \mathbf{w}^\mathsf{T}_{\mathrm{R},i} \tilde{\mathbf{y}}_i[k,m]$.
\subsection{Bistatic Range-Doppler Maps} \label{sec:periodogram}
Exploiting \ac{OFDM} signals' properties, target parameters are estimated using periodogram-based frequency estimation. The process starts with reciprocal filtering, which eliminates the transmitted symbols' influence from the beamformed received symbols by computing $g_{i,k,m} = y_{i,k,m}/x_{k,m}$ \cite{braun2010maximum}.\footnote{This work assumes that the transmitted symbols $x_{k,m}$ are known at the \acp{Rx}, either through correct demodulation or because they are a known pilot sequence, typically used for channel estimation.
}
The resulting complex value $g_{i,k,m}$ contains two sinusoids per each scatterer $\ell$, with frequencies corresponding $f_{\mathrm{d},\ell,i}$ and $\tau_{\ell,i}$.
Thus, periodogram estimation enables
the computation of a bistatic range-Doppler map for each \ac{Rx}, as follows \cite{braun2010maximum}
\begin{equation}\label{eq:period}
    \mathcal{P}_i (l, p) = \frac{1}{KM}\left|\sum_{k=0}^{K_\mathrm{p}-1} \biggl( \sum_{m=0}^{M_\mathrm{p}-1} g_{i,k,m} w_{\mathrm{K},m} e^{-j2\pi \frac{mp}{M_\mathrm{p}}}\biggr)e^{j2\pi \frac{kl}{K_\mathrm{p}}}\right|^2
\end{equation}
In \eqref{eq:period}, $w_{\mathrm{K},m}$ is the $m$th sample of an $[M \times 1]$ Kaiser window $\mathbf{w}_\mathrm{K}$ with parameter $\beta = 3$, normalized as $w_{\mathrm{K},m} = w_{\mathrm{K},m}/\sqrt{\frac{1}{M}\sum_{m=0}^{M-1}w^2_{\mathrm{K},m}}$ to preserve the statistical properties of the noise, used to reduce sidelobes in the Doppler dimension. Furthermore, $l=0,\dots, K_\mathrm{p}-1$ and $p=0,\dots, M_\mathrm{p}-1$, where $K_\mathrm{p}\geq K$ and $M_\mathrm{p}\geq M$ mean implicitly zero-padding before \ac{FFT} and \ac{IFFT} calculation. Given the constraint $R_{\mathrm{bis},i}^\mathrm{max}$ introduced in Section~\ref{sec:sysmodel}, the target search range is restricted to $l = 0, \dots, K_{\mathrm{p},i}-1$, with $K_{\mathrm{p},i} = \lfloor R_{\mathrm{bis},i}^\mathrm{max} K_{\mathrm{p}} \Delta f / c \rfloor$.

Note that a peculiar issue for bistatic sensing configurations is the \emph{blind zone}, a region near the baseline where target detection is challenging \cite{ChiGioPao:J18, Pucci:22Bistatic}. Due to bistatic range resolution, $\Delta R_\mathrm{{bis}} = \frac{c}{K_\mathrm{p} \Delta f}$, targets within a bistatic range smaller than $L_i + \Delta R_\mathrm{{bis}}$ cannot be resolved. Therefore, for reliable detection, the bistatic range must satisfy $R_\mathrm{bis} \geq L_i + \Delta R_\mathrm{{bis}}$. As a result, the bistatic range periodogram bins for the $i$th \ac{Rx} are further restricted to $l' = K'_{\mathrm{p},i}, \dots,K_{\mathrm{p},i}-1$, where $K'_{\mathrm{p},i} = \ceil*{L_i K_{\mathrm{p}}\Delta f/c}$, ensuring that periodogram bins evaluate targets' bistatic ranges at discrete values.
%
To suppress static clutter, Doppler frequencies within $[-p_\mathrm{0} / (T_\mathrm{s} M_\mathrm{p}), p_\mathrm{0} / (T_\mathrm{s} M_\mathrm{p})]$ are filtered out, where $p_\mathrm{0}$ defines the Doppler removal region.
Lastly, the bistatic range-Doppler maps computed through \eqref{eq:period} are used to derive range-angle maps trough a scan of the environment, as will be detailed in the next section.
\section{Search Operation}\label{sec:search}
This section examines the search phase, a critical stage in sensing systems where targets are first detected and localized. It occurs during network startup and periodically to update the number and positions of targets which is typically time-varying. This information is then passed to a tracking stage, which is not considered here for brevity.
\subsection{Range-Angle Map Calculation}\label{sec:range-angle}
While the \ac{Tx} illuminates the entire area of interest, each \ac{Rx} collects echoes from potential targets and generates a 
range-angle map by scanning the surrounding environment through digital beamforming. Let $N_\mathrm{dir}$ denote the total sensing directions required for a complete scan, and let $\boldsymbol{\theta}_{\mathrm{s},i} = [\theta_{1,i}, \theta_{2,i}, \dots, \theta_{N_\mathrm{dir},i}]$ represent the sensing directions for the $i$th \ac{Rx}, where the $j$th direction is given by $\theta_{j,i} = \theta_{0,i} + j\Delta \theta_\mathrm{s}$, 
%
%
with $\theta_{0,i}$ as the starting scan direction and $\Delta \theta_\mathrm{s}$ as the scan angle step. Beamforming in each direction $\theta_{j,i}$ is performed using the weight vector $\mathbf{w}_\mathrm{R}$, typically calculated as
$\tilde{\mathbf{w}}_\mathrm{R} = \mathbf{b}^\mathsf{c}(\theta_{j,i})/\sqrt{N_\mathrm{R}}$ to maximize power in the desired direction. To mitigate high sidelobes, which can degrade sensing, the array aperture is windowed by applying element-wise multiplication between $\tilde{\mathbf{w}}_\mathrm{R}$ and the window weight vector $\mathbf{c}$ \cite{Friedlander:12}. The final beamforming vector is given by ${\mathbf{w}}_\mathrm{R} = \tilde{\mathbf{w}}_\mathrm{R} \odot \mathbf{c}$,
where $\mathbf{c}$ is normalized~to ensure $\|{\mathbf{w}}_\mathrm{R} \|^2_2=1$. Here, a Dolph-Chebyshev~window~with a $30$~dB peak-to-sidelobe ratio is employed.

Since the \ac{Tx} illuminates the entire area with a wide radiation pattern, each \ac{Rx} can leverage digital beamforming to process all sensing directions from the same received signal. In particular, the $i$th \ac{Rx} applies beamforming sequentially to the received signal $\mathbf{\tilde{y}}_i[k,m]$ for each sensing direction $j$, using the corresponding weight vector $\mathbf{w}_\mathrm{R}$. The full scan duration, $T_\text{scan}$, is therefore determined by the total \ac{OFDM} signal duration, i.e., $T_\text{scan} = M \cdot T_\mathrm{s}$. Once all sensing directions are processed and each \ac{BS} has transmitted, the search mode is complete. The total search phase time is $T_\text{search} = T_\text{scan} \,|\mathcal{S}|$.\footnote{Without loss of generality, in this work, all frequency resources are allocated to sensing during the search phase, with different time slots used to ensure orthogonality between sensing and communication.}

After beamforming at the $i$th \ac{Rx}, a bistatic range-Doppler map is computed for each direction using \eqref{eq:period}.
Assuming a relatively small beamwidth, only one target is likely present in each sensing direction. Therefore, the column of the periodogram corresponding to the peak value is extracted to obtain a vector $\mathbf{r}_{j,i}=(P_{j,i}(1),\dots,P_{j,i} (K_{\mathrm{p},i}-K'_{\mathrm{p},i}))^\mathsf{T}$, representing the bistatic range profile in the $j$th direction, with
\begin{equation}
    P_{j,i}(l') = \mathcal{P}^j_{i} (l', p) |_{p = \hat{p}} \quad \text{for} \quad j = 1, \dots, N_\mathrm{dir}
\end{equation}
where $\hat{p} = \operatorname*{arg\,max}_{p} \{\mathcal{P}^j_{i}(l',p)\}$ identifies the location of the peak (i.e., the column) in the periodogram.
Repeating this process for all sensing directions yields a bistatic range-angle map $\mathbf{R}_{i} =[\mathbf{r}_{1,i},\dots, \mathbf{r}_{N_\mathrm{dir},i}]$, with $\mathbf{R}_i\in \mathbb{R}^{(K_{\mathrm{p},i}-K'_{\mathrm{p},i}) \times N_\mathrm{dir}}$. The elements of $\mathbf{R}_i$ represent the intensity at the $i$th \ac{Rx}, corresponding to a $(l', j)$ point in the bistatic range-angle domain. Each point on the map is associated with a bistatic range value, determined by the bistatic range resolution introduced earlier as $R_{\mathrm{bis},l'} = l' \Delta R_\mathrm{bis}$. 

To ensure that the maps produced by different \acp{BS} are mutually consistent, i.e., they refer to the same reference frame, a domain transformation is required. First, the bistatic range $R_{\mathrm{bis},l'}$ is converted into the target-to-\ac{Rx} range $R_{\mathrm{R},l'}$ following the approach in \cite{willis05}
\begin{equation}
\begin{aligned}
\label{eq:R2} 
R_{\mathrm{R},l'} 
 = \frac{(l' \Delta R_{\mathrm{bis}})^2 - L_i^2}{2 (l' \Delta R_{\mathrm{bis}} + L_i \sin(\theta_{j,i} - \pi/2))}.
\end{aligned}
\end{equation}

Next, for a given target-\ac{Rx} distance $R_{\mathrm{R},l'}$ and sensing direction $\theta_{j,i}$, the corresponding Cartesian coordinates, relative to a reference system centered at \ac{Rx} $i$, are computed as $(x_i, y_i) = (R_{\mathrm{R},l'} \cos{\theta_{j,i}}, R_{\mathrm{R},l'} \sin{\theta_{j,i}})$. Finally, these local $(x_i, y_i)$ coordinates must be transformed into a global $(x, y)$ reference system shared among all \acp{BS} (see, e.g., \cite{PucGio2024PEB}).

A pixel-wise pre-processing step is performed via  a binary hypothesis test to distinguish noise from potential target pixels: 
\begin{equation}\label{eq:th_test}
    \mathbf{R}_i(l', j) \overset{\mathcal{H}_1}{\underset{\mathcal{H}_0}{\gtrless}} \eta
\end{equation}
where $\mathcal{H}_0$ and $\mathcal{H}_1$ represent the noise-only and signal presence hypotheses, respectively. 
The threshold $\eta = -\sigma_\mathrm{N}^2 \ln P_\mathrm{FA,point}$ is chosen to meet the desired false-alarm probability $P_\mathrm{FA,point}$, related to the total \ac{FAR} in the search space of size $|\mathbf{R}_i|$, i.e., $\text{FAR}=P_\mathrm{FA,point}|\mathbf{R}_i|$.
This hypothesis test is applied pixel by pixel to the range-angle maps at each individual \ac{BS}, filtering out noise and retaining only potential target pixels. The final decision on target presence is made after fusion and clustering, as in Section~\ref{sec:fusioncluster}. 

\subsection{Reliability Maps for Selective Data Fusion}
When constructing range-angle maps, the finite mainlobe width at the \ac{Rx} causes the target to spread across multiple angular bins, degrading angular resolution. Similarly, as can be observed by inspecting \eqref{eq:R2}, range resolution deteriorates for small bistatic ranges. The combined effect of angular and range resolution worsens when the target is near the baseline, where even small errors in the bistatic range and \ac{DoA} can lead to significant localization errors. This aligns with the geometric dilution of precision (GDOP) of our bistatic setup.
As a result, each \ac{Tx}-\ac{Rx} bistatic pair exhibits a position-dependent target resolution, which manifests as position-dependent target smearing in the range-angle maps. In particular, pronounced target spreading is observed in specific parts of the map, especially when the target is close to the baseline. However, this issue can be leveraged to preemptively identify regions where a specific bistatic pair may be unreliable, allowing for its exclusion from data fusion.

\begin{figure}[t]
\captionsetup{font=footnotesize,labelfont=footnotesize}
    \centering
    \subfloat[][\scriptsize $\mathbf{M}_1$ -- Rx in $(-60,0)\,$m \label{fig:1}]
    {\includegraphics[width=0.48\columnwidth]{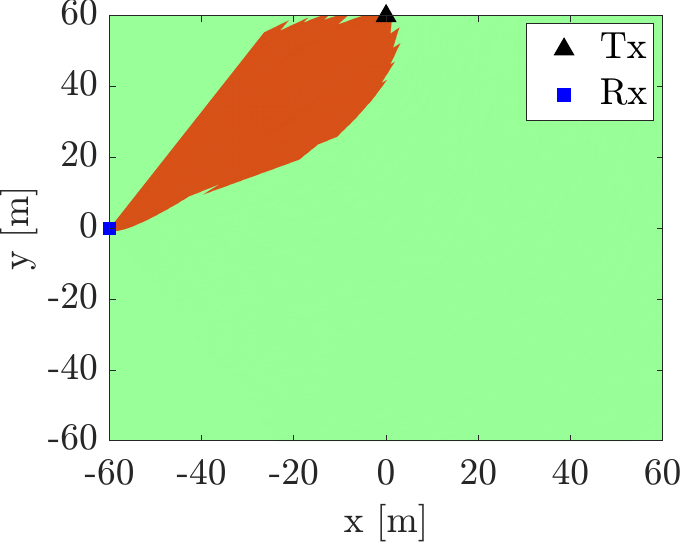}} \quad
    \subfloat[][\scriptsize $\mathbf{M}_2$ -- Rx in $(-30,-52)\,$m \label{fig:2}]
    {\includegraphics[width=0.48\columnwidth]{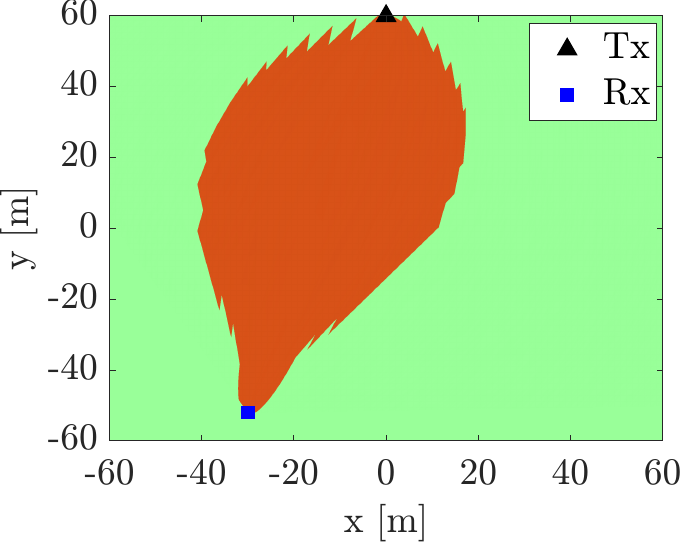}}\\ 
\caption{Example of reliability maps $\mathbf{M}_i$ for $i=1,2$ with \ac{Tx} at $(0,60)\,$m and $\gamma_\mathrm{res} = 5\,\text{m}^2$. Orange and green pixels denote $0$ and $1$, respectively.}
\label{fig:reliability_maps}
\end{figure}

To address this, we introduce reliability maps $\mathbf{M}_i\in \mathbb{R}^{(K_{\mathrm{p},i}-K'_{\mathrm{p},i}) \times N_\mathrm{dir}}$, which quantify these distortions based on the system's geometry as well as angle and range resolution, enabling the selection of portions of the range-angle maps suitable for data fusion.
For a generic  $R_{\mathrm{bis},l'}$ and \ac{DoA} $\theta_{j,i}$, four adjacent points are identified. Each point is located at $(x_i',y_i')=R_{\mathrm{R},l'}\cdot \bigl( \cos{\bigl(\theta_{j,i} \pm \delta\theta/2\bigr)}, \sin{\bigl(\theta_{j,i} \pm \delta\theta/2\bigr)}\bigr)$, where the corresponding distance $R_\mathrm{R}$ is computed using \eqref{eq:R2} by substituting $R_{\mathrm{bis},l'} \pm\frac{\Delta R_{\mathrm{bis}}}{2}$ and $\theta_{j,i} \pm \frac{\delta\theta}{2}$, and $\delta\theta$ is the beamwidth at the \ac{Rx} beamformer. 
These four points form a polygon in the Cartesian coordinate system, whose area $A_\mathrm{res}$ represents the uncertainty associated with the spatial resolution of the bistatic pair.
%
%
A threshold, $\gamma_\mathrm{res}$, is defined, above which the corresponding point in the range-angle map $\mathbf{R}_i$ is considered unreliable and, therefore, uninformative regarding the target's position. Specifically, $[{\mathbf{M}}_i]_{l', j}$ equals $1$ if $A_\mathrm{res}<\gamma_\mathrm{res}$, and $0$ otherwise.
An example of reliability maps for two bistatic pairs in the network is reported in Fig.~\ref{fig:reliability_maps}.
%
%
The reliability map is then applied to the range-angle map of each \ac{Rx} to mask unreliable points, i.e., $\overline{\mathbf{R}}_i = \mathbf{R}_i \odot \mathbf{M}_i$ for $i=1,\dots,|\mathcal{S}|-1$.
%
Selecting the appropriate area threshold is crucial: a low $\gamma_\mathrm{res}$ may remove too many regions, risking target loss, while a high $\gamma_\mathrm{res}$ preserves targets but reduces the advantage of fusing only the most informative regions for better performance.

\subsection{Fusion of Range-Angle Maps and Clustering}\label{sec:fusioncluster}
Upon completing a scan in a given network configuration $\mathcal{S}_t$, the system transitions to a new configuration in which another \ac{BS} assumes the \ac{Tx} role in a round-robin manner, i.e., $\mathcal{S}_1\rightarrow \mathcal{S}_2 \rightarrow \dots \rightarrow\mathcal{S}_{|{\mathcal S}|}\rightarrow\mathcal{S}_1\dots$. This approach enables a multi-view perception of targets by leveraging the network’s cooperative capabilities and exploiting spatial diversity, which is particularly advantageous for detecting and estimating multiple extended targets, as shadowing effects may obscure certain targets when illuminated from specific directions.

The area of interest is then divided into $R_\mathrm{x} \times R_\mathrm{y}$ grid cells, each of size $\delta_\mathrm{x} \times \delta_\mathrm{y}$, indexed by ($g_\mathrm{x}, g_\mathrm{y}$). Each range-angle map $\overline{\mathbf{R}}_i$ is therefore resampled using linear interpolation, and the resulting  $|\mathcal{S}|-1$ resampled maps from a given multistatic configuration $\mathcal{S}_t$, denoted $\boldsymbol{\Pi}_i \in \mathbb{R}^{R_\mathrm{x} \times R_\mathrm{y}}$, will be collected by the \ac{FC}. In turn the \ac{FC} fuses the maps to create a  \emph{soft map} $\boldsymbol{\Psi}_t = \sum_{i=1}^{|\mathcal{S}|-1}\boldsymbol{\Pi}_i$.%
\footnote{Note that because the range-angle maps are formed via periodogram estimation, such fusion is non-coherent. 
When complexity is not an issue coherent processing can be performed \cite{Tagliaferri:J24}.}
%
%
After computing the soft map from $|\mathcal{S}|$ different viewpoints, the \ac{FC} integrates the information into an \emph{aggregated map} given by
\begin{equation}
    \overline{\boldsymbol{\Psi}} = \sum_{t=1}^{|\mathcal{S}|} \boldsymbol{\Psi}_t\,.
    \label{eq:aggregated_map}
\end{equation}

The aggregated map can then be processed by the clustering algorithm for multiple target detection and estimation.
First, an excision filter is applied to $\overline{\boldsymbol{\Psi}}$ to further remove noise and clutter using a scenario-dependent threshold, $\gamma$. The latter is set as a fixed percentage $\gamma_\mathrm{d}$ of the peak value in the considered aggregated map, as follows
\begin{equation}
    \gamma = \gamma_\mathrm{d} \cdot \underset{(g_x, g_y)}{\max} \left\{[\overline{\boldsymbol{\Psi}}]_{g_x, g_y}\right\}.
\end{equation}
%


Next, \ac{DBSCAN} is employed to cluster regions of the map that likely correspond to targets. The \ac{DBSCAN} parameters are $\xi_\mathrm{d}$ (maximum distance) and $N_\mathrm{d}$ (minimum points per cluster) \cite{EstKriXia:96}. To enable weighted clustering based on map intensity, points exceeding the threshold are duplicated in proportion to their intensity level. This process ensures that higher-intensity points have a greater influence on \ac{DBSCAN}.
Finally, target estimation is performed by computing the weighted mean of each cluster, using the amplitude in the aggregated map of the corresponding points. The resulting position estimates are collected in the set $\hat{\boldsymbol{\mathcal{Z}}}=\{\mathbf{\hat{z}}_1,\mathbf{\hat{z}}_2,\dots, \mathbf{\hat{z}}_{|\boldsymbol{\hat{\mathcal{Z}}}|}\}$, representing the target detections extracted from the aggregated map.

\begin{table} [t]
\centering
 \resizebox{0.95\columnwidth}{!}{
 \begin{tabular}{l c | l c}
\toprule
$f_\mathrm{c}$ [GHz] & $28$ & Beamwidth $\delta\theta$ [deg] & $2.4$\\
$\Delta f$ [kHz] & $120$ & Number of BS $|\mathcal{S}|$ & $5$\\
Active subcarriers $K$ & $3168$ & $T_\mathrm{scan}$ [ms] & $3$\\
OFDM symbols $M$ & $336$ & $\theta_0$ [deg] & $-58.8$\\
$N_\mathrm{T}$, $N_\mathrm{R}$ & $50$ & Area threshold $\gamma_\mathrm{res}$ $[\text{m}^2]$ & $5$\\
$N_\mathrm{dir}$ & $50$ & FAR & $10^{-2}$\\
\bottomrule
\end{tabular}}
\caption{JSC System and Scenario Parameters}
\label{tab:param}
\end{table}
\section{Numerical Results}\label{sec:results}
Key parameters shared by all \acp{BS} are listed in Table~\ref{tab:param}. Each \ac{Rx} collects $M = 336$ \ac{OFDM} symbols modulated with a \acs{QPSK} alphabet. The average \ac{RCS} can be either $\bar{\sigma}_\mathrm{rcs}=0.5\,$m$^2$ or $\bar{\sigma}_\mathrm{rcs}=5\,$m$^2$. The \ac{PSD} is $N_0 = k_\mathrm{B} T_0 F$, where $T_0 = 290\,\text{K}$, $F = 13\,\text{dB}$, and $k_\mathrm{B} = 1.38 \times 10^{-23}\,\text{JK}^{-1}$. The system monitors a $120\,\text{m} \times 120\,\text{m}$ area using $|\mathcal{S}| = 5$ \acp{BS}, located at ($0\,\text{m}, 60\,\text{m}$), ($-60\,\text{m},0\,\text{m}$), ($-30\,\text{m},-52\,\text{m}$), ($30\,\text{m},-52\,\text{m}$), and ($60\,\text{m},0\,\text{m}$). Target locations are uniformly distributed within a $70\,\text{m} \times 70\,\text{m}$ area centered at $(0,0)$, and are independently drawn during $N_\mathrm{MC} = 1000$ Monte Carlo iterations ensuring a minimum separation of $1\,\text{m}$ to avoid overlapping. Each target moves with velocity $\mathbf{v} = (v_\mathrm{x}, v_\mathrm{y})$, where $v_\mathrm{x}, v_\mathrm{y}$ are drawn uniformly from $[-20\,\text{m/s}, 20\,\text{m/s}]$. Clutter is modeled using $\mathcal{L} = 25$ static point-like scatterers, which have null Doppler and are uniformly distributed within the surveillance area.  
%
\begin{figure}[t]
    \centering
    \includegraphics[width=0.65\columnwidth]{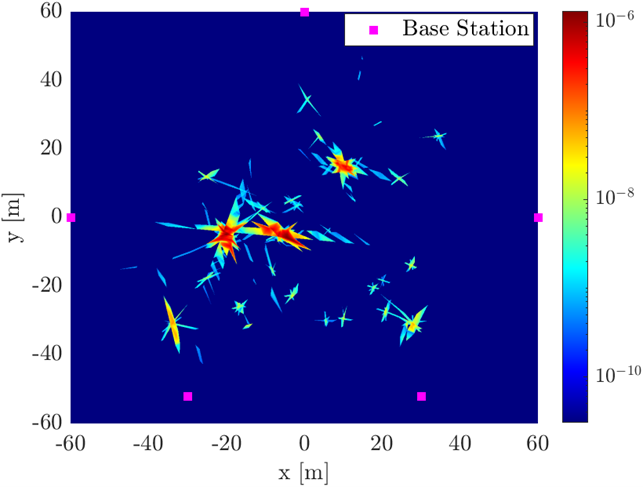}
    \caption{Example of aggregated map with $Q=3$ targets computed according to \eqref{eq:aggregated_map} where each \ac{BS} is a \ac{Tx} in a round-robin fashion. Targets are at positions $(-20\,\text{m}, -5\,\text{m})$, $(-6\,\text{m}, -5\,\text{m})$, and $(10\,\text{m}, 15\,\text{m})$, with respective velocities of $(17,19)\,\text{m/s}$, $(19,13)\,\text{m/s}$, and $(3,2)\,\text{m/s}$.}
    \label{fig:Aggregated_map}
\end{figure}
During clustering a threshold $\gamma_\mathrm{d} = 5\%$ is considered, with $\xi_\mathrm{d} = 2$ and $N_\mathrm{d} = 50$. The resampled maps $\mathbf{\Pi}$ generated by each \ac{BS} in receiving mode cover the entire surveillance area, with $R_\mathrm{x} = R_\mathrm{y} = 701$ grid points and a resolution of $\delta_\mathrm{x} = \delta_\mathrm{y} = 0.1\,\text{m}$. An example of an aggregated map, constructed using \eqref{eq:aggregated_map}, is shown in Fig.~\ref{fig:Aggregated_map}.

Given the complex scenario involving multiple targets, the \ac{GOSPA} metric is employed to assess the sensing performance of the network. 
The $p$-order \ac{GOSPA} metric, computed for each Monte Carlo iteration, is defined as \cite{RahGarSve:17}  
\begin{equation}\label{eq:GOSPAdef}
        \mathrm{GOSPA}\! =\!
        \Bigg[\frac{1}{N_\mathrm{c}}\Bigg(\sum_{(i,j)\in\boldsymbol{\zeta}_\mathrm{g}^*}\hspace{-8pt}\|\mathbf{z}_i - \mathbf{\hat{z}}_j\|_p^p+\frac{\xi_\mathrm{g}^p}{2}(|\boldsymbol{\mathcal{Z}}|+ |\boldsymbol{\hat{\mathcal{Z}}}|-2|\boldsymbol{\zeta}_\mathrm{g}^*|)\Bigg)\Bigg]^{\!\!\frac{1}{p}}
\end{equation}
where $\boldsymbol{\hat{\mathcal{Z}}}$ is the set of estimated target positions, and $\xi_\mathrm{g}$ is the gating parameter. This parameter ensures that any estimate with a potential assignment distance greater than $\xi_\mathrm{g}$ is treated as a false detection, while unmatched true positions are considered missed detections. Moreover, the vector $\boldsymbol{\zeta}_\mathrm{g}^*$ represents the optimal assignment—i.e., the assignment that minimizes the metric—between estimated and true targets. Finally, the total number of terms in the summation is given by $N_\mathrm{c} = |\boldsymbol{\hat{\mathcal{Z}}}|+ |\boldsymbol{\mathcal{Z}}|-|\boldsymbol{\zeta}_\mathrm{g}^*|$.
For $p=2$, the first term in \eqref{eq:GOSPAdef} represents the mean squared error, while the second term defines detection rate, $R_\mathrm{D} = |\boldsymbol{\zeta}_\mathrm{g}^*|/|\boldsymbol{\mathcal{Z}}|$, false detection rate, $R_\mathrm{FA}=(|\boldsymbol{\hat{\mathcal{Z}}}|- |\boldsymbol{\zeta}_\mathrm{g}^*|)/|\boldsymbol{\hat{\mathcal{Z}}}|$, and missed detection rate, $R_\mathrm{MD}=1-R_\mathrm{d}$.
%
%
In this study, a \ac{GOSPA} order of $p = 2$ is adopted, with a gating parameter of $\xi_\mathrm{g} = 5\,\text{m}$.

\begin{figure*}[t]
    \centering
    \includegraphics[width=0.95\linewidth]{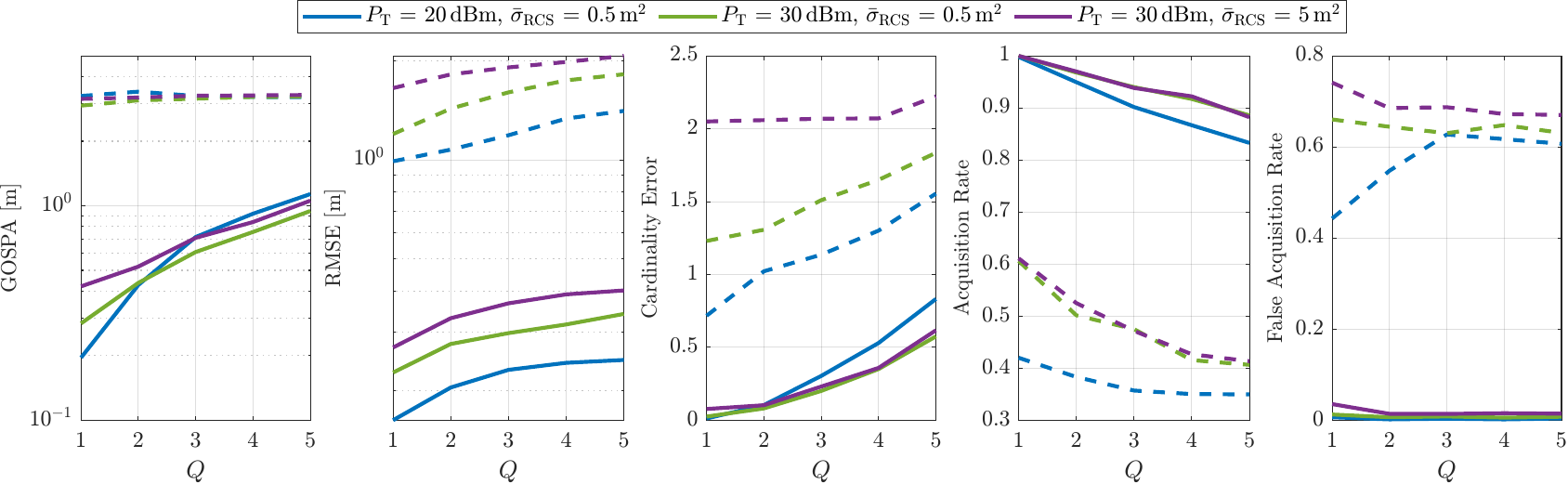}
    \caption{Results in the search phase showing the \ac{GOSPA}, \ac{RMSE}, cardinality error, acquisition rate, and false acquisition rate as a function of the number of targets present in the surveilled area, obtained by averaging over the Monte Carlo iterations. Dashed lines represent performance when the reliability maps $\mathbf{M}$ are not applied.}
    \label{fig:GOSPA_search}
\end{figure*}


Fig.~\ref{fig:GOSPA_search} shows system performance as a function of the number of targets. The results indicate that higher transmitted power leads to increased \ac{GOSPA} error if $Q \leq 2$, prompting the use of \ac{RMSE} to measure localization accuracy. The \ac{RMSE} that is computed for correctly assigned targets as $\text{RMSE} = \sqrt{(\sum_{q=0}^{|\boldsymbol{\zeta}_{\mathrm{g}}|-1} (\zeta_{\mathrm{g},q}-\zeta_{\mathrm{g},q}^*)^2)/|\boldsymbol{\zeta}_{\mathrm{g}}|}$, where $\boldsymbol{\zeta}_{\mathrm{g}} \subseteq \boldsymbol{\mathcal{Z}}$ is the subset of actual targets that are optimally assigned in $\boldsymbol{\zeta}_{\mathrm{g}}^*$, provides a clearer localization metric. 
%
%
An unexpected trend is observed: performance degrades as $P_\mathrm{T}$ increases. This counterintuitive behavior can be attributed to the nature of the aggregated map, which becomes less clear due to residual clutter, distortions due to the interpolation step used for map alignment consistency, and smeared target representations. This last effect is caused by a misalignment between the actual target \ac{AoA} and the scanning directions. Indeed, especially near the baseline, adjacent beams at the \ac{Rx} tend to overlap, causing targets to appear elongated in the aggregated map. As explained, this issue is mitigated via reliability maps, which select only those portions of the maps that do not exhibit target smearing before their fusion. Nonetheless, higher $P_\mathrm{T}$ can still exacerbate map contamination, leading to a deterioration in performance.

Additionally, results show a degradation in \ac{GOSPA} performance for $P_\mathrm{T} = 20\,$dBm, primarily due to an increase in cardinality errors. This degradation arises from two factors: the noise threshold $\eta$ used to filter each range-angle map and the \ac{DBSCAN} clustering parameters. A high noise threshold may result in smaller clusters that the clustering algorithm fails to identify as targets, classifying them as outliers. Fig.~\ref{fig:GOSPA_search} illustrates that this degradation is linked to a lower acquisition rate, indicating that missed acquisitions, rather than false alarms, are the primary cause. 
Notably, with a well-designed cooperating system, the localization error can be kept below $40\,$cm even when the transmit power and \ac{RCS} are relatively low, i.e., $P_\mathrm{T}=20\,$dBm and $\bar{\sigma}_\mathrm{rcs} = 0.5\,\text{m}^2$.

\section{Conclusion}\label{sec:conclu}
We investigated a multistatic \ac{MIMO}-\ac{OFDM} system that leverages the cooperation of multiple \acp{BS} for detecting and localizing targets. We proposed a selective fusion strategy that filters range-angle maps based on a pre-calculated distortion metric, which accounts for bistatic geometry. Additionally, we introduced a round-robin transmitter role strategy among \acp{BS} to enable multi-view target perception.
Using the \ac{GOSPA} and \ac{RMSE} metrics, we demonstrated that, counterintuitively, increasing $P_\mathrm{T}$ led to higher \ac{GOSPA} errors, primarily due to map contamination from residual clutter and target smearing caused by bistatic configurations. In this context, we showed that reliability maps effectively mitigate these impairments, significantly reducing both \ac{GOSPA} and \ac{RMSE} by a factor of five in scenarios with three targets.


\bibliographystyle{./Bibliography/IEEEtran}
\bibliography{./Bibliography/IEEEabrv,./Bibliography/bibliography}

\end{document}